\documentclass[reprint,amsmath,amssymb,aps,prb]{revtex4-2}
\usepackage{graphicx}
\usepackage[percent]{overpic}
\usepackage[version=3]{mhchem}
\usepackage{graphicx}
\usepackage{dcolumn}
\usepackage{bm}
\usepackage{hyperref}
\hypersetup{
    colorlinks=true,
    linkcolor=blue,
    filecolor=blue,
    urlcolor=blue,
   citecolor=blue,
}
\usepackage{enumitem}
\usepackage{color}
\usepackage{amsmath}
\usepackage{comment}
\usepackage{siunitx}
\usepackage{bm}
\usepackage{chemformula}
\usepackage{tikz-feynman}

\begin{document}
\title{Cavity-enhanced superconductivity via band engineering}

\author{Valerii K. Kozin}
\affiliation{Department of Physics, University of Basel, Klingelbergstrasse 82, CH-4056 Basel, Switzerland}
\author{Even Thingstad}
\affiliation{Department of Physics, University of Basel, Klingelbergstrasse 82, CH-4056 Basel, Switzerland}
\author{Daniel Loss}
\affiliation{Department of Physics, University of Basel, Klingelbergstrasse 82, CH-4056 Basel, Switzerland}
\author{Jelena Klinovaja}
\affiliation{Department of Physics, University of Basel, Klingelbergstrasse 82, CH-4056 Basel, Switzerland}

\date{\today}

\begin{abstract}
We consider a two-dimensional electron gas interacting with a quantized cavity mode. We find that the coupling between the electrons and the photons in the cavity enhances the superconducting gap. Crucially, all terms in the Peierls phase are kept, in contrast to more naive approaches, which may result in spurious superradiant phase transitions. We use a mean-field theory to show that the gap increases approximately linearly with the cavity coupling strength. The effect can be observed locally as an increase in the gap size via scanning tunneling microscopy (STM) measurements for a flake of a 2D material (or for a Moir\'e system where the enhancement is expected to be more pronounced due to a large lattice constant) interacting with a locally-structured electromagnetic field formed by split-ring resonators. Our results are also relevant for quantum optics setups with cold atoms interacting with the cavity mode, where the lattice geometry and system parameters can be tuned in a vast range. 
\end{abstract}

\maketitle

\section{Introduction} 
Placing condensed matter systems in an optical cavity is a promising way to engineer their properties through quantum fluctuations of the cavity field~\cite{cavQuantMatReview, Hbener2020, Appugliese2022}. 
For instance, it has been shown that one may tune electron systems across (topological) phase transitions through coupling to cavity modes~\cite{Dmytruk2022_ControllingTopologicalPhases, PhysRevX.10.041027,PhysRevB.99.235156, PhysRevResearch.6.033188}. Cavity coupling can also be used to tune the magnetic interaction in Mott insulators, which may be essential to realize strongly fluctuating  magnetic phases depending on the delicate balance between competing interactions~\cite{Zou2024_Dissipative, Thingstad2024_FractionalSpinQuantum}.
For 1D electron systems embedded in a cavity~\cite{PhysRevB.108.085410,PhysRevB.107.045425,PhysRevLett.131.023601}, it has been predicted that charge density wave (CDW) order is enhanced at the expense of superconducting order at half filling~\cite{PhysRevLett.125.217402}. This is very typical behavior in low-dimensional systems, where CDW is often strongly favoured due to the very simple Fermi surface. Moreover, the development of long range order is very often suppressed in one-dimensional systems due to strong fluctuations. Yet, many of the most interesting and useful phenomena in condensed matter physics are based precisely on the development of long range orders such as superconductivity, magnetism, and charge ordering. Understanding and exploiting the full potential of cavities as a technique to manipulate condensed matter systems therefore requires moving from one-dimensional to higher-dimensional systems. Motivated by this, as a first step, band structure engineering has recently been very actively pursued in non-interacting systems ~\cite{welakuh2023nonperturbative,PhysRevResearch.4.013012}.

Intriguing many-body effects that were recently predicted are cavity-induced~\cite{SchlawinAmpereanSC, DrivingPairing,DrivingFluctuationsPairing,PhysRevB.105.165121} and cavity-engineered~\cite{Sentef2018,PhysRevLett.122.167002,PhysRevB.99.020504, PhysRevResearch.2.013143} (topological~\cite{Ricco2022,Ricco2022a,PhysRevB.107.115418}) superconductivity. In the paper by Schlawin \textit{et al.}~\cite{SchlawinAmpereanSC}, it was argued that a two-dimensional electron gas (2DEG) placed in a cavity undergoes a {cavity induced} superconducting phase transition at a reasonably high critical temperature $\sim10^0-10^1$~ K. However, such large numbers could only be obtained by boosting the light-matter coupling strength using the argument of cavity-volume compression, which is applicable only for the sub-wavelength cavities ~\cite{PhysRevB.109.104513}. Instead, it applies for split ring resonators, which are LC circuits with resonance frequency $\omega_0=1/\sqrt{LC}$, where $L$ and $C$ are the inductance and capacitance of the lumped elements of the effective circuit. The electric field is uniform and almost completely localized in the capacitor, whose size can be substantially smaller than the cavity mode wavelength $\lambda_\text{cav}=2\pi c_0/(\sqrt{\varepsilon} \omega_0)$, where $c_0$ is the speed of light and ${\varepsilon}$ the relative permittivity of the material surrounding the resonator. The effective mode volume can be written in the form~\cite{andberger2023terahertz} $V_{\text{eff}}=A_{\text{mv}} (\lambda_\text{cav}/2)^3$, where $A_\text{mv}$ is the mode volume compression factor. Since the effective light-matter coupling strength is proportional to $1/\sqrt{V_\mathrm{eff}}$, it can be strongly enhanced by  decreasing the effective mode volume. The mode volume compression factors can reach values of $<10^{-5}$~\cite{ModeCompressionPRB} or even $<10^{-10}$~\cite{ModeCompressionNanoLett}, thus offering a promising platform for substantial modification of the material properties via the strong light-matter interaction. Furthermore, it was recently shown experimentally that extremely high mode volume compression can be achieved with microcubes uniformly distributed over the surface of a 2D material~\cite{Epstein2020_FarFieldExcitation}. The main difference from conventional Fabry--P\'{e}rot-type cavities~\cite{MicrocavitiesKavokin} is that the modes of split-ring resonators (or other resonators that support high mode volume compression) are quasi-static and also do not carry well-specified momentum.

In this paper, we revisit the effect of light-matter coupling on the superconducting instability  and switch from Fabry--P\'{e}rot cavities hosting transverse modes to the case of quasi-static modes of split-ring resonators where the mode volume can be tuned in a vast range to allow for strong light-matter coupling. Considering an electronic hopping Hamiltonian coupled to the cavity modes through Peierls substitution, we use mean-field theory to show that the cavity modes renormalize the electron spectrum. This enhances the density of states for the electron system (akin to mechanical strain or pressure~\cite{Locquet1998,Ruf2021,Wang2022,Engelmann2013}), which may in turn enhance superconductivity. {Instead of considering dominant forward scattering, as in Ref.~\cite{Sentef2018}, we consider superconductivity induced by an attractive onsite Hubbard interaction.} We further suggest that this can be measured experimentally as a local enhancement of the superconducting gap in a superconductor partially covered by a split-ring resonator. This opens new avenues to control and to manipulate the properties of superconductors  
in a symbiotic relationship with electromagnetic environments.

\section{Model} 
\begin{figure}
    \centering
    \includegraphics[width=0.85\columnwidth]{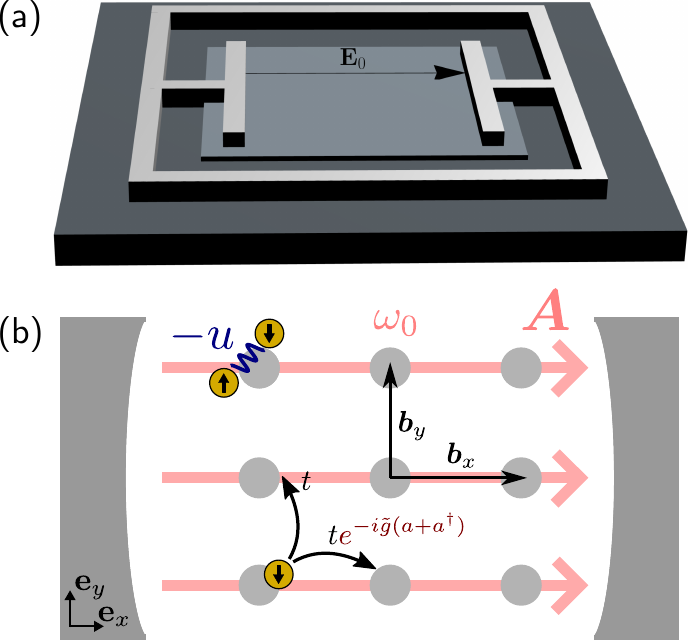}
    \caption{(a) The sketch of the system shows a 2D supercondcuting electron system (the thin plate on a substrate) enclosed in a single mode cavity, implemented by, e.g., a split-ring resonator (an LC-circuit). The arrow shows the direction of the electric-field oscillations. {(b) A simplified figure showing the tight-binding description of the 2D system.} 
    }
    \label{fig_1}
\end{figure}

We consider a two-dimensional superconductor, consisting of an electron gas with attractive electron-electron interaction, that is coupled to a cavity such as a split-ring resonator (see Fig.~\ref{fig_1}). The cavity modes are quasi-static and can be described by the scalar potential $\mathbf{E}(\mathbf{r},z,t)=-\nabla \phi(\mathbf{r},z,t)$. Using a gauge transformation, one may alternatively describe such a field only in terms of an electromagnetic vector potential $\mathbf{A}$. We assume that the geometry of the resonator is chosen such that the corresponding vector potential is oscillating along the unit vector $\mathbf{e}_x$ along the $x$-axis, so that we have $\mathbf{A} = \sqrt{\hbar /(2\epsilon_0 \epsilon \omega_0 V_\mathrm{eff})} \mathbf{e}_x (a + a^\dagger)$, where $\epsilon_0$ is the vacuum permittivity, $\epsilon$ the relative permittivity, $V_\mathrm{eff}$ the effective mode volume, and $a^\dagger$ and $a$ are the bosonic creation and annihilation operators, respectively, for the photonic cavity mode. 

We further describe the electron system by a tight binding model with $t_j$ being the hopping amplitude along lattice translation vectors $\mathbf{b}_j$ of a Bravais lattice. We assume that the effect of the electromagnetic field on the electrons is captured by Peierls substitution, so that the system is modelled by 
\begin{align}\label{eq_H}
    &H=-\sum_{\mathbf{i},\sigma, j} (t_j e^{-{ie\mathbf{A}\cdot\mathbf{b}_j}/{\hbar}}c^\dagger_{\mathbf{i}+\mathbf{b}_j,\sigma}c_{\mathbf{i},\sigma}+\text{h.c.} ) \nonumber\\
    &-u\sum_{\mathbf{i}}n_{\mathbf{i},\uparrow}n_{\mathbf{i},\downarrow}+\hbar\omega_0 a^\dagger a,
\end{align}
where $c_{\mathbf{i},\sigma}$ and $c^\dagger_{\mathbf{i},\sigma}$ are annihilation and creation operators for electrons on lattice site $\mathbf{i}$ with spin $\sigma$ (and here $n_{\mathbf{i},\sigma}=c^\dagger_{\mathbf{i},\sigma}c_{\mathbf{i},\sigma}$),  $e$ is the electron charge, $u>0$ is the amplitude of the attractive interaction. Since we assumed a position-independent (uniform) vector potential, the Peierls phase 
can be rewritten as $\tilde{g}_{j}(a+a^\dagger)$ with effective coupling strength $\tilde{g}_{j} \equiv \sqrt{W_{j}/\omega_0}$ depending on the hopping direction, where 
\begin{equation}
    W_j=\frac{1}{\hbar}\frac{e^2 (\mathbf{e}_x\cdot\mathbf{b}_j)^2}{2\varepsilon_0\varepsilon V_\text{eff}}.
\end{equation} 
For simplicity, we consider a square lattice with lattice constant $a_0$ and  assume that the vector potential is oriented along a lattice translation vector, say $\mathbf{b}_x$, in the following. To simplify the notation, we therefore let $\tilde{g} \equiv \tilde{g}_x$ and have $\tilde{g}_y = 0$. 

\section{Mean-field theory} 
We employ a mean-field decoupling of the hopping term by assuming that the ground state wave function $|GS\rangle$ is a direct product of matter and cavity states $|\psi_m^\mathrm{el} \rangle$ and $|\psi_l^\mathrm{ph} \rangle$ through $|GS\rangle=|\psi_m^\mathrm{el} \rangle|\psi_l^\mathrm{ph}\rangle$. Furthermore, we also introduce a standard BCS mean-field decoupling of the electron-electron interaction to obtain mean-field equations for the coupled system. {The product-state approximation for the ground state becomes exact for non-interacting electron systems placed in a single-mode cavitiy~\cite{PhysRevResearch.4.013012,Eckhardt2022}.}
Other interactions may lead to light-matter entanglement in finite systems~\cite{PhysRevLett.131.023601,PhysRevResearch.6.033188}, {which, however, may still lead to the same values of ground-state observables in the (quasi-) thermodynamic limit~\cite{PhysRevResearch.6.033188,PhysRevLett.125.143603}.}

After the mean-field decoupling, we arrive at the  Hamiltonian \(H = E_0 + H_\mathrm{el}^\mathrm{MF} + H_\mathrm{ph}^\mathrm{MF} \), where the electronic and photonic mean-field Hamiltonians, respectively, are given by 
\begin{align}
H_\mathrm{el}^\mathrm{MF} &= \sum_{\mathbf{k} \sigma} \xi_{\mathbf{k}} c_{\mathbf{k}\sigma}^\dagger c_{\mathbf{k}\sigma} + \sum_{\mathbf{k}} (\Delta c_{\mathbf{k} \uparrow}^\dagger c_{-\mathbf{k} \downarrow}^\dagger + h.c. ), \\
H_\mathrm{ph}^\mathrm{MF} &=\hbar\omega_0 a^\dagger a +  C \cos [\tilde{g} (a + a^\dagger) ], \label{eq_mfHPhotonic}  
\end{align}
 where we introduced Fourier transformed electron creation and annihilation operators $c_{\mathbf{k}}^\dagger$ and $c_{\mathbf{k}}$, the constant energy shift $E_0 = - \gamma C + \Delta^2 / u$,
and the renormalized bare electron spectrum  and mean-field parameters defined as
\begin{align}
\xi_{\mathbf{k}}& = - 2 t \gamma \cos (k_x a_0) - 2 t \cos (k_y a_0) - \mu,\\
C &=  -2 t \sum_{\mathbf{k}\sigma}  \cos (k_x a_0) \langle c_{\mathbf{k}\sigma}^\dagger c_{\mathbf{k}\sigma} \rangle, \\
\Delta &= -u \sum_{\mathbf{k}} \langle c_{-\mathbf{k} \downarrow} c_{\mathbf{k} \uparrow} \rangle,\\
\gamma &= \langle \cos [\tilde{g} (a + a^\dagger) ] \rangle \label{eq_mfGamma},
\end{align}
where $\mu$ is the chemical potential and $t_j \equiv t$ for simplicity. Here, we assume that the electron distribution function \(\langle  c_{\mathbf{k},\sigma}^\dagger c_{\mathbf{k},\sigma} \rangle\) remains symmetric with respect to \(\mathbf{k} \rightarrow -\mathbf{k}\). As discussed in App.~\ref{app_mfDecoupling}, this assumption can be relaxed to obtain a more general mean-field Hamiltonian allowing for spontaneous breaking of this inversion symmetry. However, we have shown that the associated mean-field parameter describing this asymmetry relax to zero, so that the somewhat simplified mean-field description above describes the ground state equally well.  

We can now derive the mean-field equations and get
\begin{align} 
C &=
\sum_{\mathbf{k}} (-2 t \cos k_x a_0) \left[ \frac{E_{\mathbf{k}}-{\xi}_{\mathbf{k}}}{2E_{\mathbf{k}}}  +  \frac{{\xi}_\mathbf{k}n_F(E_\mathbf{k})}{E_\mathbf{k}}  \right],  \label{eq_mfC}\\
\Delta &= \frac{1}{N}\sum_{\mathbf{k}} \frac{\Delta}{2 E_{\mathbf{k}}} \tanh \left( \frac{\beta E_{\mathbf{k}}}{2} \right)\label{eq_mfDelta}, 
\end{align} 

\noindent  with the spectrum \(E_{\mathbf{k}} = \sqrt{{\xi}_{\mathbf{k}}^2 + |\Delta|^2 }\), where \(\beta^{-1} = k_{\text{B}} T\) is the inverse temperature, $N$ the number of lattice sites, and $n_F$ the Fermi-Dirac distribution. The expectation value in Eq.~\eqref{eq_mfGamma} is calculated with respect to the mean-field Hamiltonian in Eq.~\eqref{eq_mfHPhotonic}, and depends on the electronic mean-field parameter \(C\). On the other hand, the right-hand sides of Eqs. \eqref{eq_mfC} and \eqref{eq_mfDelta} for \(C\) and \(\Delta\) depend on \(\gamma\) through the electronic spectrum. 

The mean-field parameters can be determined iteratively. 
The self-consistent expression for \(\gamma\) can be determined numerically. When there are many electrons in the system, however, one may derive an approximate analytical expression. In that limit, we have \(|C| \gg \hbar\omega_0\), so that the second term in \(H_\mathrm{ph}^\mathrm{MF}\) dominates. Thus, the expectation value of and the fluctuations in \(\tilde{g} ( a + a^\dagger)\) are small. This allows a second-order Taylor expansion of the cosine function. The mean-field Hamiltonian becomes then quadratic and can be diagonalized exactly (see App.~\ref{app_diagonalizationPhHamiltonian}). The associated ground state is 
\begin{equation}
     |\psi^\mathrm{ph}_0 \rangle = \mathcal{N} e^{\frac{1}{2}[\xi^* a^2-\xi (a^\dagger)^2]}|0\rangle,
\end{equation}
where $|0\rangle$ is the photon vacuum state,  and \(\mathcal{N}\) is a normalization constant. Furthermore, $\xi=(1/2)\ln{(\tilde{\omega}/\omega_0})$ with $\tilde{\omega}=\sqrt{\omega_0^2+\omega_p^2}$, where \(\omega_p \equiv \tilde{g}  (2|C|  \omega_0/\hbar)^{1/2}\) is the plasma frequency. For the mean-field parameter \(\gamma\), we then find 
\begin{align}
\gamma \simeq 1 - \frac{\tilde{g}^2}{2} \frac{ \omega_0 }{\sqrt{\omega_0^2 + \omega_p^2} },
\label{eq_mfGammaV2}
\end{align}
\noindent 
{which coincides with the result from Ref.~\cite{Eckhardt2022}. 
Importantly, this band squeezing should not be confused with the effective mass renormalization in Refs.~~\cite{welakuh2023nonperturbative,PhysRevResearch.4.013012}, which is due to photons dressing  single-particle excitations.} When \(\gamma\) is close to 1, the fluctuations in \(\tilde{g} (a + a^\dagger)\) are necessarily small, and the expansion above is valid. We can then solve the mean-field Eqs.~(\ref{eq_mfC}), (\ref{eq_mfDelta}), and (\ref{eq_mfGammaV2}). 

In the following, we solve the gap equation [see Eq.~\eqref{eq_mfDelta}] at zero temperature to determine how the coupling to the cavity modifies the superconducting gap $\Delta$. {First, we provide an approximate solution by assuming that \(\delta \gamma \equiv 1 - \gamma \ll  1\).} Subsequently, we solve the mean-field equations numerically. 
Within the mean-field description, the only effect of coupling the electron system to the cavity is squeezing of the electronic band, an effect which is analogous to dynamical localization ~\cite{Eckhardt2022, welakuh2023nonperturbative,PhysRevResearch.4.013012,PhysRevB.105.165121}. We now discuss how this affects the amplitude of the superconducting gap. 

In the BCS regime the zero-temperature gap amplitude is of the form
$\Delta=2\hbar\omega_c e^{-{1}/{\lambda}}$,
where $\omega_c$ is an effective cut-off frequency and the effective coupling strength $\lambda$ is proportional to the density-of-states $\nu$ and the amplitude of the effective interaction mediating the pairing, i.e. \(u\). 
The former is inversely proportional to the hopping amplitude, which results in $\nu\propto1/\sqrt{\gamma}$. If the cavity only introduces weak band squeezing, this gives
$\nu\approx\nu_0(1+\delta\gamma/2)$, where $\nu_0$ is the density-of-states in the absence of cavity coupling. When the effective coupling strength in the absence of cavity coupling is $\lambda_0$, this gives $\lambda=\lambda_0(1+\delta\gamma/2)$ and 
\begin{equation} \label{eq_schematicSol}
\Delta 
\approx \Delta_0 \exp \left( \frac{\delta\gamma}{2\lambda_0} \right),
\end{equation}
where $\Delta_0 = 2\hbar\omega_c e^{-1/\lambda_0}$ is the gap in the absence of coupling to the cavity~\footnote{If we have two modes, then there is no 1/2 factor in front of $\delta \gamma$ because in this case the band is squeezed in both directions.}. Since \(\lambda_0\) is typically small, the renormalization of the electronic band structure can be small and still give rise to significant renormalization of the superconducting gap.

{The formula for the plasma frequency can be simplified in the effective-mass approximation.} We consider a sample of area \(l^2\) with electron density \(n\) and electrons described by a parabolic dispersion relation with band mass \(m\). Effectively, this system can be described by the hopping Hamiltonian in Eq.~\eqref{eq_H}, where the hopping amplitude is \(t = \hbar^2/(2 m a_0^2)\). Assuming that the band renormalization is small so that \(|C| \simeq 2 N_\mathrm{el} |t| \), where $N_\mathrm{el}$ is the number of free electrons in the system, the plasma frequency occurring in the mean-field equations is given by~\cite{PhysRevResearch.4.013012} 
\begin{equation}
    \omega_p=\sqrt{\frac{e^2 n l^2}{\varepsilon_0\varepsilon m V_\text{eff}}}.
\end{equation}
{We also rederive this expression diagrammatically in  App.~\ref{section_app_diag_deriv}.}

{We now estimate the renormalization of the superconducting gap}. For 2D materials such as TMDs, typical parameters~\cite{Geim2013} are  $n=0.5\times 10^{11}$~cm$^{-2}$, $l=2$~$\SI{}{\micro\meter}$, and $m=0.5m_{e}$. This gives Fermi energy $E_F=0.25$~meV and $N_\text{el}\approx2000$. With $\varepsilon A_\text{mv}=10^{-10}$, $\varepsilon=12$, and $\omega_0=2\pi \times 8$~THz we obtain a band renormalization factor $\gamma\approx0.995$ ($\delta \gamma = 0.005$). Since we consider a local Hubbard interaction, we assume $\hbar\omega_c=4t$
with $t=\SI{156} {\milli\electronvolt}$,
and with a bare gap  $\Delta_0 = 0.05$~meV, this corresponds to an effective coupling strength $\lambda_0 \approx 0.1$. From Eq.~\eqref{eq_schematicSol}, the resulting gap amplitude renormalization is $\delta \Delta/\Delta_0 \approx 0.03$. This gap renormalization is small, but significant compared to the level spacing \(\delta E=1/(\nu_{2D}l^2)=\hbar^2\pi/(m l^2) \), as $\delta \Delta/\delta E\approx 10$. Thus, the effect should be observable. 

In Eq.~\eqref{eq_mfGammaV2}, it is clear that since \(\omega_p \propto \sqrt{N_\mathrm{el} t}\), the band renormalization is the largest for systems with small hopping parameter $t$, corresponding to systems with small bandwidth, and for systems with a small number of electrons.  Both features are indeed present in twisted bilayer systems, which have recently been actively studied in cavities~\cite{PhysRevLett.131.176602,masuki2024cavity,PhysRevLett.132.166901}, and are also known for hosting superconducting phases~\cite{Cao2018,Bistritzer2011}. {Also, such a flat-band dispersion can be engineered in a 2DEG by applying gates~\cite{Krix2024}.} Thus, we may reasonably expect
that these systems are promising candidates to host large band structure renormalization, and correspondingly large renormalization of the superconducting gap. A very naive estimate can be obtained by considering systems which instead have a large lattice constant $a_0 = \SI{10}{\nano\meter}$~\cite{PhysRevB.105.205424,Andrei2021,Kennes2021,Dean2013,Wang2015,Forsythe2018} and a small effective mass $m=0.1 m_e$. The second order expression in Eq.~\eqref{eq_mfGammaV2} then gives 
$\gamma = 0.93$, which shows that a significant band structure renormalization should be possible. With such a value of $\gamma$, taking the cavity frequency $\omega_0=2\pi \times 3$~THz, and density $n= 10^{11}$~cm$^{-2}$ gives 
$E_F = \SI{2.4}{\milli\electronvolt}$ and $N_\mathrm{el} \approx 4000$. 
Assuming a bare gap $\Delta_0\approx0.1$~meV and the cut-off $\hbar\omega_c=15$~meV, the simple estimate above gives $\lambda_0 \approx 0.2$ and substantial enhancement $\delta\Delta/\Delta_0\approx0.23$, while $\delta \Delta / \delta E \approx 40$. {Importantly, we performed all the calculations at zero temperature, which is justified as long as $k_B T\ll k_B T_c\ll \hbar \omega_0$ which corresponds to the parameter range considered above.} {For non-ideal resonators, the equilibrium ground state properties remain qualitatively the same as long as the cavity decay rate is relatively small~\cite{PhysRevResearch.6.033188}.}

\section{Numerical results} 
\begin{figure}
    \centering
    \includegraphics[width=0.98\columnwidth]{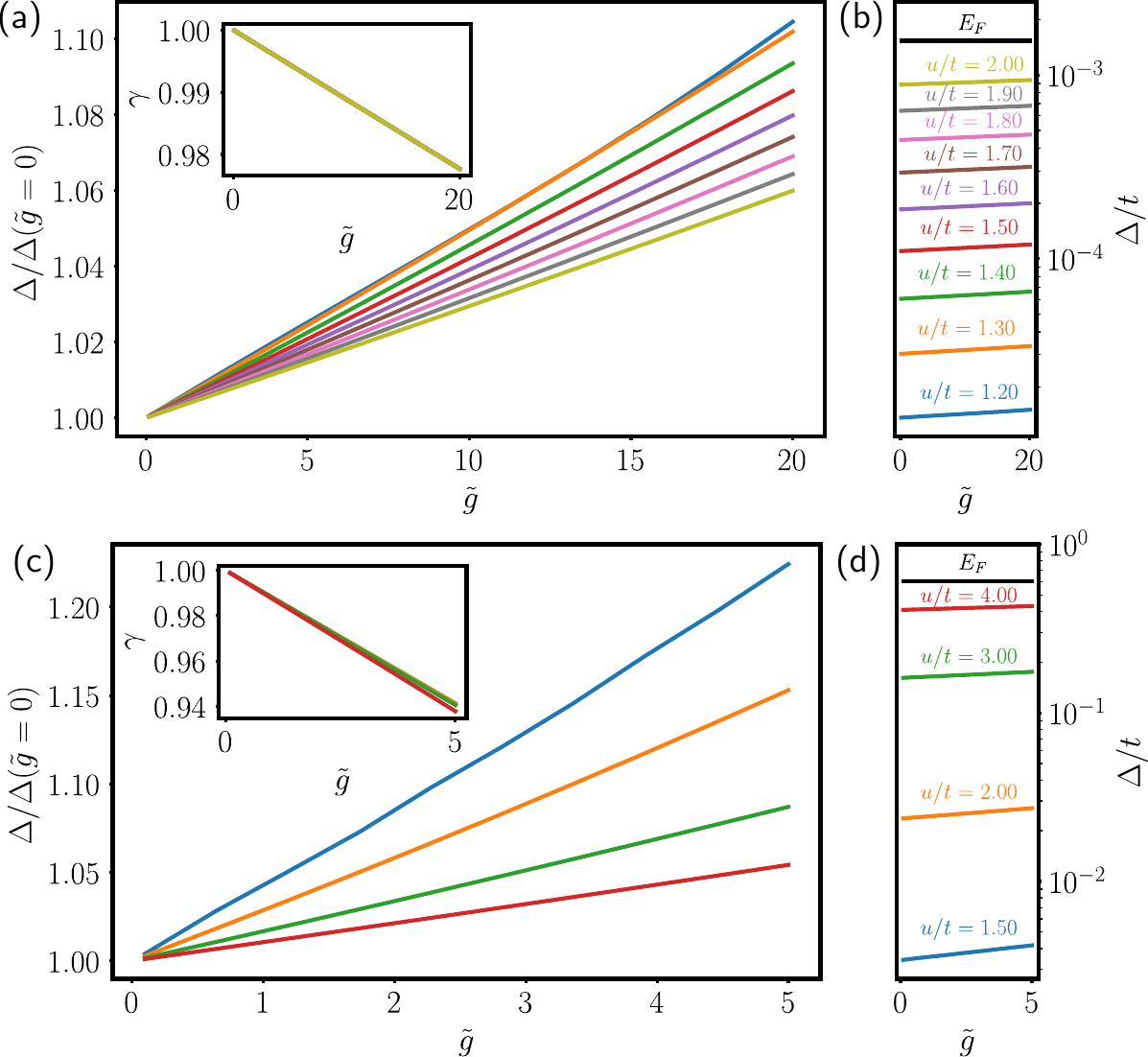}
    \caption{Numerical results:  
   Gap $\Delta$ as  function of cavity coupling strength \(\tilde{g}\) (normalized to the value in the absence of coupling)  (a)  for a single quantum well (or a 2D material) and (c) for the case of a large lattice constant and small effective mass, corresponding to an effective model for electrons on a Moir\'e lattice. The insets show the band structure renormalization factor \(\gamma\). With increasing cavity coupling strength the band renormalization becomes larger and the Fermi surface is more squeezed. This increases the density of states, which in turn increases the gap.  (b,d) Gap size for various interaction strengths $u$ in units of $t$. The remormalization of $\Delta$ is the strongest at small $u/t$.}
    \label{fig_numRes}
\end{figure}

Now we solve the self-consistent equations, 
Eqs.~\eqref{eq_mfGamma}-\eqref{eq_mfDelta}, 
numerically using the parameters given above. 
The result is shown in Fig.~\ref{fig_numRes}(a), which shows the superconducting gap $\Delta$ at zero temperature as function of the cavity coupling strength \(\tilde{g}\) for a number of different interaction ratios \(u/t\). As shown in the inset, the band structure renormalization is of the same order as predicted by the simplified analysis in the previous subsection. We find that the gap amplitude increases linearly with the cavity coupling strength. Since the band renormalization strength $\delta \gamma$  is small compared to the effective coupling strength $\lambda_0$ for the given interaction strengths, this is in accordance with what we expect from Eq.~\eqref{eq_schematicSol}. For the physically relevant case of having gaps smaller than the Fermi energy, the band renormalization factor $\delta \gamma$ depends very weakly on the development of a superconducting gap. We also calculate the cavity enhancement of the superconducting gap for larger lattice constants and smaller effective masses, to obtain an order of magnitude estimate of the effect in twisted bilayer systems, see Fig.~\ref{fig_numRes}(b).

\section{Discussion} The physical reason behind the cavity-induced enhancement is dynamical localization or the polaron effect~\cite{mahan1990many,Eckhardt2022}:
the electrons get dressed by a cloud of virtual photons {which suppress the hopping amplitude. This enhances the density of states in 2D, and consequently also the superconducting gap. 
The effect only appears beyond the effective mass approximation. 
}
Fundamentally, the reason for the relatively small enhancement of the gap is similar to why there is no superradiance~\cite{PoliniNoGo1}. Since the number of electrons is large compared to the number of cavity modes, the electron system can have a strong impact on the cavity mode, while the inverse effect is much weaker. Above, we considered coupling to a single cavity mode. The effect can be further enhanced by including many modes~\cite{Eckhardt2022}, which would amount to $\tilde{g}\to\tilde{g}\sqrt{N_\text{modes}}$, or by applying driving~\cite{Eckhardt2024,DrivingPairing}.
By considering only a single mode with vector potential aligned in a particular direction, the Fermi surface was stretched in that direction. 
We assume that superconductivity is induced by an on-site attractive interaction, and this gives rise to an isotropic gap, even if the Fermi surface itself is slightly anisotropic. With more sophisticated pairing interactions, we would expect anisotropic gap enhancement. 
By instead considering multiple cavity modes with  modes oscillating in orthogonal directions~\cite{andberger2023terahertz}, the Fermi surface would instead exhibit isotropic stretching and larger gap enhancement as pointed out above. We note that Ref.~\cite{PhysRevLett.125.217402} investigated intertwined orders in such a system and found that superconducting correlations were suppressed at the expense of charge density wave correlations. However, charge density wave order benefits strongly from the very simple Fermi surface in one dimension, and there is no reason to expect similar behaviour in higher dimensional systems.  Additionally, we point out that in Ref.~\cite{Sentef2018}, the effect of a cavity on the phonon interaction was shown not to lead to an enhancement of the superconducting gap in a setup involving a Fabry--P\'{e}rot resonator.
In experiments, the proposed effect could be observed as a local increase in the gap via STM-measurement for a flake of $\mathrm{MoS_2}$ (or a superconducting twisted bilayer system) with split-ring resonator on top for a subregion~\cite{andberger2023terahertz} of the sample. Our results are also relevant for quantum optics setups with cold atoms/ions interacting with the cavity mode~\cite{Schlawin2019_CavityMediatedUnconventional,PhysRevLett.111.185301,PhysRevLett.111.185302}.

\section{Conclusion} In this study, we have investigated the interplay between a two-dimensional electron gas with pairing interaction and a quantized cavity mode, unveiling a mechanism to enhance the superconducting gap through interaction with cavity photons. 
We consider an superconducting electron system described by a Hubbard model with attractive interaction which couples to the cavity through a Peierls phase. Utilizing a mean-field theory, we demonstrate a linear augmentation of the superconducting gap amplitude with increasing cavity coupling strength. The effect could be observed through a local enhancement of the superconducting gap through STM measurements in 2D materials or even Moir\'e systems, where the enhancement is notably accentuated by the large lattice constants and large density of states. Our work opens new avenues for manipulating superconductivity through cavity quantum electrodynamics and lays the foundation for future exploration of the symbiotic relationship between electronic systems and electromagnetic environments.

\section{Acknowledgments} We thank Dante M. Kennes for useful discussions. The work of VKK was supported by the Georg H. Endress Foundation. 
This work was supported  by the Swiss National Science Foundation. 
This project received funding from the European Union’s Horizon 2020 research and innovation program (ERC Starting Grant, Grant Agreement No. 757725).
 
\appendix
\section{Mean-field decoupling}\label{app_mfDecoupling}

In momentum space, the Hamiltonian $H$ from Eq.~(\ref{eq_H}) in the main text takes the form

\begin{align}
H = \hbar\omega_0 a^\dagger a + \sum_{\mathbf{k}\sigma} \hat{\xi}_{\mathbf{k}} c_{\mathbf{k}\sigma}^\dagger c_{\mathbf{k}\sigma} 
- u \sum_{\mathbf{k}\mathbf{k}'\mathbf{q}} c_{\mathbf{k}+\mathbf{q}\uparrow}^\dagger c_{\mathbf{k}'-\mathbf{q} \downarrow}^\dagger c_{\mathbf{k}'\downarrow} c_{\mathbf{k} \uparrow} ,
\end{align}
\noindent where  \(\hat{\xi}_\mathbf{k}\) is the operator is defined as
\begin{align}
\hat{\xi}_\mathbf{k} =  - 2 t \cos [k_x a_0 - \tilde{g} (a + a^\dagger)] - 2 t \cos k_y a_0 - \mu. 
\end{align}
Through a mean-field decoupling of the hopping term, we obtain the mean-field Hamiltonian \(H= E_0 + H_\mathrm{ph}^\mathrm{MF} + H_\mathrm{el}^\mathrm{MF}\), where the constant energy shift is $E_0 = - \gamma C-\phi S + \Delta^2 / u$ and the photonic mean-field Hamiltonian is given by
\begin{align}
H_\mathrm{ph}^\mathrm{MF}= \hbar\omega_0 a^\dagger a + C \cos [ \tilde{g} (a + a^\dagger)]   + S \sin [\tilde{g} ( a +a^\dagger)].
\end{align}
\noindent The electronic mean-field Hamiltonian in the absence of the gap is 
\begin{align}
H^{(0)}_\mathrm{el} =  \sum_\mathbf{k} \xi_\mathbf{k} c_{\mathbf{k}\sigma}^\dagger c_{\mathbf{k} \sigma},
\end{align}
\noindent with the mean-field spectrum
\begin{align}
\xi_{\mathbf{k}} = - 2 t \gamma \cos{k_x a_0} - 2 t \phi \sin{k_x a_0} - 2 t \cos{k_y a_0} - \mu,
\end{align}
\noindent where we have introduced mean-field parameters 
\begin{align}
C &= \sum_{\mathbf{k} \sigma} (-2 t \cos{k_x a_0}) \langle c_{\mathbf{k}\sigma}^\dagger c_{\mathbf{\mathbf{k}}\sigma} \rangle, \label{eq_C_appendix}\\
S &= \sum_{\mathbf{k}\sigma} (-2 t \sin{k_x a_0}) \langle c_{\mathbf{k}\sigma}^\dagger c_{\mathbf{k}\sigma} \rangle, \label{eq_S_appendix}\\
\gamma &= \langle \cos [ \tilde{g} (a +a^\dagger)] \rangle, \\
\phi &= \langle \sin [\tilde{g} (a + a^\dagger)] \rangle.
\end{align}
The spectrum can now also be written in the form 
\begin{align}
\xi_\mathbf{k} &= -2 t \sqrt{\gamma^2 + \phi^2 } \cos [ k_x a_0 - \arg (\gamma + i \phi) ] \nonumber\\
&- 2 t \cos k_y a_0 - \mu.
\end{align}
The dispersion relation therefore has a minimum at \((k_x, k_y) = (\arg (\gamma + i \phi), 0)/a_0 \equiv \mathbf{Q}\), and is shifted away from \((0,0)\) if the system develops a finite expectation value for \(\phi\). 

Next, we now perform a mean-field decoupling also of the attractive electron-electron interaction term in the superconducting channel. From the phase space arguments, this should happen at net momentum \(\mathbf{Q}\). We therefore introduce the 
mean-field parameter 
\begin{align}
b_{\mathbf{Q},\mathbf{p}} = \langle c_{\mathbf{Q}-\mathbf{p},  \downarrow} c_{\mathbf{Q} + \mathbf{p}, \uparrow} \rangle .
\end{align}
Furthermore, we introduce the gap as
\begin{align}
\Delta_\mathbf{Q} = - u \sum_\mathbf{p} b_{\mathbf{Q},\mathbf{p}}.
\end{align}
We therefore consider the BCS Hamiltonian
\begin{align}
&H_\mathrm{el}^\mathrm{MF} = \sum_{\mathbf{k}\sigma} \xi_{\mathbf{k}} c_{\mathbf{k}\sigma}^\dagger c_{\mathbf{k}\sigma}  \nonumber\\
&+ \sum_{\mathbf{p}} \left( \Delta_{\mathbf{Q}} c_{\mathbf{Q}+\mathbf{p}\uparrow}^\dagger c_{\mathbf{Q}-\mathbf{p} \downarrow}^\dagger + \Delta^*_{\mathbf{Q}} c_{\mathbf{Q}-\mathbf{p} \downarrow} c_{\mathbf{Q}+\mathbf{p} \uparrow} \right).
\end{align}
We now introduce the momentum variable \(\mathbf{p} = \mathbf{k} - \mathbf{Q}\). The electronic Hamiltonian then takes the form 
\begin{align}
H_\mathrm{el}^\mathrm{MF}= \sum_\mathbf{p} 
\begin{pmatrix} c_{\mathbf{Q}+\mathbf{p} \uparrow}^\dagger c_{\mathbf{Q}-\mathbf{p} \downarrow}  \end{pmatrix}
\begin{pmatrix}
\tilde{\xi}_\mathbf{p} & \Delta_\mathbf{Q} \\ \Delta_\mathbf{Q}^* & -\tilde{\xi}_{-\mathbf{p}} 
\end{pmatrix}
\begin{pmatrix} c_{\mathbf{Q}+\mathbf{p} \uparrow} \\ c^\dagger_{\mathbf{Q}-\mathbf{p} \downarrow} \end{pmatrix},
\end{align}
\noindent where \(\tilde{\xi}_\mathbf{p} \equiv \xi_{\mathbf{p} + \mathbf{Q}}\) is the shifted spectrum. 

This Hamiltonian can be diagonalized through a unitary transformation. Effectively, the coupling to the photons therefore renormalizes the hopping amplitude \(t\). We may now obtain the gap equation for the gap \(\Delta_\mathbf{Q}\) by deriving its self-consistent equations or minimizing the free energy. This results in 
\begin{align}
\Delta_\mathbf{Q} = \frac{1}{N}\sum_\mathbf{p} \frac{\Delta_\mathbf{Q}}{2 E_\mathbf{p}} \tanh \left( \frac{\beta E_{\mathbf{p}}}{2} \right),
\label{eq_selfConsistentDeltaQ}
\end{align}
\noindent with the spectrum \(E_\mathbf{p} = \sqrt{\tilde{\xi}_\mathbf{p}^2 + |\Delta_\mathbf{Q}|^2 }\). From the unitary transformation utilized to diagonalize the Hamiltonian, we obtain
\begin{align}
\langle c_{\mathbf{Q}+\mathbf{p} \uparrow}^\dagger c_{\mathbf{Q}+\mathbf{p} \uparrow} \rangle = \langle c_{\mathbf{Q}-\mathbf{p},\downarrow}^\dagger c_{\mathbf{Q}-\mathbf{p}, \downarrow} \rangle = \eta_\mathbf{p}
\end{align}
with 
\begin{align}
    \eta_\mathbf{p} = \frac{1}{2}\left( 1 - \frac{\tilde{\xi}_\mathbf{p}}{E_\mathbf{p}}  \right) + \frac{\tilde{\xi}_\mathbf{p} }{E_\mathbf{p}}  n_F(E_\mathbf{p}) .
\end{align}

The electron distribution function $n_{\mathbf{p},\uparrow}=\eta_{\mathbf{p}-\mathbf{Q}}$ and $n_{\mathbf{p},\downarrow}=\eta_{\mathbf{p}-\mathbf{Q}}$ and, thus, when $\mathbf{Q}$ is zero we have $n_{\mathbf{p},\sigma}=n_{-\mathbf{p},\sigma}$.
Shifting the momentum in Eq.~(\ref{eq_C_appendix}) and Eq.~(\ref{eq_S_appendix}) by $\mathbf{Q}$ and exploiting $\cos{Q_x}=\gamma/\sqrt{\gamma^2+\phi^2}$ and $\sin{Q_x}=\phi/\sqrt{\gamma^2+\phi^2}$, we then obtain the mean-field equations
\begin{align}
C = \frac{\gamma}{\sqrt{\gamma^2 + \phi^2}} \sum_\mathbf{p} (-4t \cos{p_x a_0}) \eta_\mathbf{p}, \\
S = \frac{\phi}{\sqrt{\gamma^2 + \phi^2}}\sum_\mathbf{p} (-4t \cos{p_x a_0}) \eta_\mathbf{p},
\end{align}
where we used \(E_\mathbf{p} = E_{-\mathbf{p}} \). Together with Eq.~\eqref{eq_selfConsistentDeltaQ} and the mean field equations for $\gamma$ and $\phi$, these equations can now be solved iteratively.

\section{Approximate diagonalization of the mean-field photon Hamiltonian}\label{app_diagonalizationPhHamiltonian}

We start with writing down the mean-field photon Hamiltonian from the main text
\begin{align}
H_\mathrm{ph}^\mathrm{MF} = C \cos [\tilde{g} (a + a^\dagger)] + S \sin [ \tilde{g}(a+a^\dagger)] + \hbar\omega_0 a^\dagger a.
\end{align}
It can be rewritten in the following form
\begin{align}
H_\mathrm{ph}^\mathrm{MF} = -\sqrt{C^2 + S^2} \cos [\tilde{g} (a + a^\dagger) - \Phi]  + \hbar\omega_0 a^\dagger a,
\end{align}
\noindent where \(\Phi = \arg (-C - i S)\). Assuming \(A \equiv \sqrt{C^2 + S^2} \gg \hbar\omega_0\), the average value of and the fluctuations in \(\tilde{g} (a+a^\dagger) - \Phi\) should be small in the ground state. This allows a Taylor expansion of the cosine:
\begin{align}
H^{\text{MF}}_{\text{ph}} &\simeq  - \sqrt{C^2 + S^2} ( 1- \Phi^2/2) - \frac{\hbar\omega_0}{2}\nonumber\\
&- \sqrt{C^2 + S^2}\, \tilde{g} \Phi\, (a + a^\dagger)\nonumber\\ &+\frac{\hbar\omega_0}{2} \begin{pmatrix} a^\dagger & a \end{pmatrix} \begin{pmatrix} 1+\rho & \rho \\ \rho & 1+\rho \end{pmatrix} \begin{pmatrix} a \\ a^\dagger \end{pmatrix} ,
\end{align}
where we have introduced \(\rho = \tilde{g}^2 \sqrt{C^2 + S^2}  / (\hbar\omega_0) \). 
We further introduce the Bogoliubov transformation 
\begin{align}
\begin{pmatrix} a \\ a^\dagger \end{pmatrix} = \begin{pmatrix} u & -v \\ -v & u \end{pmatrix} \begin{pmatrix} \alpha \\ \alpha^\dagger \end{pmatrix}
\end{align}
\noindent with matrix elements 
\begin{align}
u = \frac{1}{\sqrt{2}} \left( \frac{1 + \rho}{\kappa} + 1 \right)^{1/2}, 
\quad
v = \frac{1}{\sqrt{2}} \left( \frac{1 + \rho}{\kappa} - 1 \right)^{1/2}, 
\end{align}
\noindent where \(\kappa = \sqrt{(1+\rho)^2 - \rho^2 } = \sqrt{1+2\rho}\). We then obtain 
\begin{align}
H = \tilde{E}_0 + \kappa \hbar\omega_0 (\alpha^\dagger - \delta) (\alpha - \delta) ,
\end{align}
\noindent where the constant energy shift is given by
\begin{align}
\tilde{E}_0 = -\sqrt{C^2 + S^2} (1 - \Phi^2/2) + \frac{\hbar\omega_0}{2} (\kappa -1) - \kappa \omega_0 \delta^2
\end{align}
and 
\begin{align}
\delta = \Phi \tilde{g} \sqrt{C^2 + S^2} / (\hbar\omega_0 \kappa^{3/2}) .
\end{align}
We may now make sure that the fluctuations in $\tilde{g} (a+a^\dagger) - \Phi$ are indeed small, so that the expansion is justified. Furthermore, we may calculate the average value
\begin{align}
\gamma = \langle \cos [ \tilde{g} (a + a^\dagger) ] \rangle \approx 1- \frac{1}{2} \tilde{g}^2 \langle (a+a^\dagger)^2 \rangle 
\end{align}
in the ground state of the photonic mean-field Hamiltonian. For the special case $S = \phi = 0$, this gives 
\begin{align}
\gamma \approx 1 - \frac{\tilde{g}^2}{2\kappa},
\end{align}
which is equivalent with the expression given in terms of the plasma frequency in the main text.

\section{Diagrammatic calculation of the dressed photon propagator}
\label{section_app_diag_deriv}
{
In this section we derive the renormalization of the cavity frequency in the effective mass (continuum) approximation for a 2D non-interacting electron gas. Let us consider a finite-sized quantum well interacting with a uniform cavity mode oscillating in the plane of the 2DEG. The system is described by the following second-quantized Hamiltonian
\begin{equation}
    H=\sum_{\sigma}\int d\mathbf{r} \psi_{\sigma}^{\dagger}(\mathbf{r})\frac{(-i\nabla+e{\mathbf{A}})^2}{2m}\psi_{\sigma}(\mathbf{r})+\omega_0 a^{\dagger}a,
    \label{eq:ham}
\end{equation}
where $\psi_\sigma(\mathbf{r})$ are the electron field operators, $\mathbf{A}=A_0\mathbf{e}_x (a+a^\dagger)$ with $A_0 = \sqrt{1/(2\epsilon\epsilon_0\omega_0 V_{\text{eff}})}$, $m$ is the effective mass, and we set $\hbar=1$.

\tikzset{
    gluon/.style={decorate, draw=black,
    decoration={snake,amplitude=4pt, segment length=5pt}}
}
\begin{figure}[h]
\[ 
\vcenter{\hbox{\begin{tikzpicture}
  \begin{feynman}
    \vertex[] (a) {};
    \vertex[right=1.5 cm of a] (b) {};
    \diagram*{
      (a) -- [photon, very thick] (b),
    };
  \end{feynman}
\end{tikzpicture}}}
=
\vcenter{\hbox{\begin{tikzpicture}
  \begin{feynman}
    \vertex[] (a) {};
    \vertex[right=1.4cm of a] (b) {};
    \diagram*{
      (a) -- [photon] (b),
    };
  \end{feynman}
\end{tikzpicture}}}
+
\vcenter{\hbox{\begin{tikzpicture}
  \begin{feynman}
    \vertex[] (a) {};
    \vertex[right=1.4cm of a, dot] (b) {};
    \vertex[right=1cm of b, dot] (c) {};
    \vertex[right=1.4cm of c] (d) {};  
    \diagram*{
      (a) -- [photon] (b),
      (b) -- [with arrow=0.5, half left, looseness=1.2] (c),
      (c) -- [fermion, half left, looseness=1.2] (b),
      (c) -- [photon,very thick] (d),
    };
  \end{feynman}
\end{tikzpicture}}}
\]
\[~+~
\hbox{\begin{tikzpicture}
  \begin{feynman}
    \vertex[] (a) {};
    \vertex[right=1.4cm of a, dot] (b) {};
    \vertex[right=1.4cm of a, dot] (b1) {};
    \vertex[right=1.4cm of b] (c) {};  
    \diagram*{
      (a) -- [photon] (b),
      (b1) -- [with arrow=0.5, loop, min distance=1.5cm] (b),
      (b1) --[photon,very thick](c),
    };
  \end{feynman}
\end{tikzpicture}}
\]

    \centering
    \caption{
{The diagrammatic representations of the Dyson equation for the cavity photon field, which incorporates two terms: the particle-hole (bubble) diagram stemming from the paramagnetic interaction term $\mathbf{p}\cdot\mathbf{A}$ and the tadpole diagram stemming from the diamagnetic term $\mathbf{A}^2$.}}
    \label{fig_diagr}
\end{figure}

We begin with calculating the dressed photon propagator, shown in Fig.~\ref{fig_diagr}. We introduce the photon propagator in the Matsubara representation ~\cite{PhysRevB.108.085410,PhysRevB.98.235123}:
\begin{align}
&D_0(\tau) = - \langle \mathcal{T} {A}_x(\tau){A}_x(0)\rangle,\nonumber\\
&D_0(i\omega) = \frac{2 A_0^2 l^2}{\omega_0}\frac{\omega_0^2}{(i\omega)^2-\omega_0^2} \, ,  \label{D0}
\end{align}
where $\tau$ is the imaginary time, $\omega$ is the bosonic Matsubara frequency, $\mathcal{T}$ stands for the time ordering operation, $D_0(i \omega)$ is the Fourier transform of $D_0(\tau)$, and $l^2$ is the area of the system which appears after performing the momentum-space Fourier transform. Since the cavity mode is uniform and thus cannot transfer any momentum, we omit the momentum argument. Coupling of the cavity mode to the 2DEG results in modification (dressing) of the photon propagator through
\begin{align}
    D(i\omega)=\left[D_0^{-1}(i\omega)-\mathcal{P}(i\omega)\right]^{-1},
\end{align}
where $\mathcal{P}(i\omega)$ is the polarization operator represented by the particle-hole (bubble) and the diamagnetic (tadpole) diagrams in Fig.~\ref{fig_diagr}. 

{The particle-hole diagram $\pi(i\omega,\mathbf{q})$ yields zero if the momentum $\mathbf{q}$ transferred by the cavity mode is vanishing: 
\begin{align}
    &\pi(i\omega,\mathbf{q}) =2\frac{e^2}{m^2}\int\frac{d\mathbf{k}}{(2\pi)^2}\int\frac{d\omega'}{2\pi}G_0(i\omega',\mathbf{k})k_x\nonumber\\
    &\times G_0(i\omega'-i\omega,\mathbf{k}-\mathbf{q})(k_x-q_x)\nonumber\\
    &=\frac{2e^2}{m^2}\int\frac{d\mathbf{k}}{(2\pi)^2}\int\frac{d\omega'}{2\pi}\frac{k_x}{i\omega'-\xi_\mathbf{k}}\frac{k_x-q_x}{i\omega'-i\omega-\xi_{\mathbf{k}-\mathbf{q}}}\nonumber\\
    &=\frac{2e^2}{m^2}\int \frac{d\mathbf{k}}{(2\pi)^2}\frac{\theta(-\xi_{\mathbf{k}+\mathbf{q}})-\theta(-\xi_\mathbf{k})}{\xi_{\mathbf{k}+\mathbf{q}}-\xi_\mathbf{k}-i\omega}k_x(k_x+q_x),
    \label{eq_particleHoleExpression}
\end{align}
where the electron Green function is given by $G_0(i\omega,\mathbf{k})=(i\omega-\xi_{\mathbf{k}})^{-1}$ with $\xi_{\mathbf{k}}=\mathbf{k}^2/(2m)-E_F$, and where $E_F$ is the Fermi energy. When $\mathbf{q}=0$, one gets $\pi(i\omega,\mathbf{q}=0)=0$, in other words, all intra-band transitions with zero transferred momentum are prohibited by the Pauli exclusion principle. }

The tadpole diagram stems from the diamagnetic term and it reads
\begin{align}
    &\mathcal{P}(i\omega)=\pi_{\text{dia}}(i\omega)=\frac{2e^2}{m} \int\frac{d\omega'}{2\pi}\frac{d\mathbf{k}}{(2\pi)^2}G_0(i\omega',\mathbf{k})\nonumber\\
    &=2\frac{e^2}{n}\int\frac{kdk}{(2\pi)^2}2\pi\theta(-\xi_\mathbf{k})=\frac{e^2 n}{m},
\end{align}
where we have one factor 2 originating from the diagrammatic rules and a second factor of 2 from the spin summation. Plugging this result into the Dyson equation we arrive at
\begin{equation}
D(i\omega) = \frac{l^2}{\epsilon_0\epsilon V_\text{eff}\omega_0^2}\frac{\omega_0^2}{(i\omega)^2-(\omega_0^2+\omega_p^2)},
\end{equation}
where $\omega_p=\sqrt{{e^2 n l^2}/{(\varepsilon_0\varepsilon m V_\text{eff})}}$ is the plasma frequency so we note that the dressing leads to the renormalization of the cavity frequency (the pole of the photon Green function) $\omega_0\to\tilde{\omega}_0=\sqrt{\omega_0^2+\omega_p^2}$.

{
The vanishing of particle-hole bubble is a consequence of treating the model system to be spatially uniform, so that only the propagator at momentum $\mathbf{q} = 0$ is relevant. In reality, the system will always have a finite extent, and this will make cavity modes associated with a corresponding momentum scale relevant as well. In the following, however, we demonstrate that the contribution from the particle-hole diagram is small as long as $v_F q \ll \omega$ (where $q$ is the magnitude of the momentum associated with a given mode and $v_F=k_F/m$ the Fermi velocity), and that this is the relevant regime in our case.

To analyze the paramagnetic contribution $\pi(i\omega,\mathbf{q})$ at finite momentum $\mathbf{q}$, we evaluate the expression in Eq.~\eqref{eq_particleHoleExpression} at finite small $q$ such that $q \ll k_F$ and $v_F q / \omega \ll 0$. We then compare with the diamagnetic contribution and show that the paramagnetic contribution is small in comparison.  
From the Heaviside step functions in the numerator, it follows that only small regions close to the Fermi surface can contribute to the integral, namely the regions where $\{k<k_F, |{\mathbf{k}+\mathbf{q}}|>k_F\}$ and $\{k>k_F, |{\mathbf{k}+\mathbf{q}}|<k_F\}$. Exploiting this together with $n=k_F^2/(2\pi)$ and using $d^2 k=k dk d\phi\approx k_F dk d\phi$, where $\phi$ is the relative angle between $\mathbf{k}$ and $\mathbf{q}$, we obtain 

\begin{align}
&\frac{\pi(i\omega,\mathbf{q})}{e^2n/m}\approx -\frac{2}{k_F}\nonumber\\
&\times\int_0^{2\pi}\frac{d\phi}{2\pi}\frac{\cos{\phi}\; \cos{(\phi+\phi_q)}[k_F\cos{(\phi + \phi_q)}+q\cos{\phi_q}]}{\cos{\phi}-\frac{i\omega}{q v_F}}
\end{align}
for small $q$, where $\phi_q$ is the angle between the $x$-axis and $\mathbf{q}$. Explicit computation up to second order in $q$ then gives 
\begin{align}
&\frac{\pi(\omega,\mathbf{q})}{e^2n/m} \simeq-\frac{1}{2} \left( \frac{v_F q}{\omega}\right)\nonumber\\
&\times\left[ \left( \frac{ v_F q}{\omega}\right) \left( \frac{1}{2} \cos 2 \phi_q +1\right) + i\frac{2 q}{k_F} \cos^2 \phi_q \right].
\end{align}
We see that as long as the characteristic transferred momentum is small compared to the characteristic transferred frequency, the paramagnetic part of the polarization operator remains small. To make an estimate for the relative size of the two contributions, we assume that the mode is slightly non-uniform, i.e. the characteristic transferred momentum is $q\sim1/l$, where $l$ is the lateral size of the sample. We further assume the characteristic frequency to be $\omega_0$, although in reality, we may expect much larger frequencies to be relevant since the renormalization $\omega_0\to\sqrt{\omega_0^2+\omega_p^2}$ is large. For the parameters taken in the main text, we then have 
\begin{equation}
    \frac{v_F q}{\omega_0}=\frac{k_F q}{m\omega_0}=\frac{\sqrt{2E_F /{ml^2}}}{\omega_0}\approx 10^{-2},
\end{equation}
which justifies the assumption of a uniform cavity mode.

}

\bibliography{bibliography}
\end{document}